\documentclass[10pt,journal,compsoc]{IEEEtran}
\ifCLASSOPTIONcompsoc
  \usepackage[nocompress]{cite}
\else
  \usepackage{cite}
\fi
\ifCLASSINFOpdf
   \usepackage[pdftex]{graphicx}
  \graphicspath{{../pdf/}{../jpeg/}}
  \DeclareGraphicsExtensions{.pdf,.jpeg,.png}
\else
  \usepackage[dvips]{graphicx}
  \graphicspath{{../eps/}}
  \DeclareGraphicsExtensions{.eps}
\fi
\usepackage{mathtools, amsmath}
\usepackage{algorithm}
\usepackage{algpseudocode}
\algnewcommand\algorithmicinput{\textbf{Input:}}
\algnewcommand\algorithmicoutput{\textbf{Output:}}
\algnewcommand\algorithmicbl{\textbf{\# of blocks:}}
\algnewcommand\algorithmicth{\textbf{\# of threads / block:}}
\algnewcommand\Input{\item[\algorithmicinput]}
\algnewcommand\Output{\item[\algorithmicoutput]}
\algnewcommand\Block{\item[\algorithmicbl]}
\algnewcommand\Thread{\item[\algorithmicth]}
\algrenewcommand\algorithmicindent{1.0em}

\usepackage{pifont}
\usepackage{tabularx}
\usepackage{array,booktabs}
\usepackage{multirow}
\usepackage{slashbox}
\usepackage{dblfloatfix}
\newcolumntype{C}{>{\centering\arraybackslash}X} 
\usepackage{lipsum}

\usepackage{soul}
\usepackage{array}
\usepackage{color}
\usepackage{nicefrac}

\usepackage{xcolor}
\usepackage{colortbl}

\newcolumntype{P}[1]{>{\centering\arraybackslash}p{#1}}

\newtheorem{Example}{Example}

\newcommand{\mTest}{I}

\newcommand{\mRefAlg}[1]{Algorithm~\ref{#1}}
\newcommand{\mRefFig}[1]{Fig.~\ref{#1}}
\newcommand{\mRefEq}[1]{Equation~\ref{#1}} 

\newcommand{\mRefSec}[1]{Section~\ref{#1}}

\usepackage{hyperref}
\begin{document}
%
\title{ParaLiNGAM: Parallel Causal Structure Learning for Linear non-Gaussian Acyclic Models}
%
%
%
%

\author{Amirhossein~Shahbazinia,
		Saber~Salehkaleybar,~
		and~Matin~Hashemi
\thanks{
All authors are with Learning and Intelligent Systems Laboratory, Department
of Electrical Engineering, Sharif University of Technology, Tehran, 
Iran. \protect \\
E-mails: amirhossein.shahbazinia@ee.sharif.edu, saleh@sharif.edu (Corresponding author), matin@sharif.edu. Webpage: http://lis.ee.sharif.edu 
}
}

\IEEEtitleabstractindextext{%
\begin{abstract}
One of the key objectives in many fields in machine learning is to discover causal relationships among a set of variables from observational data. In linear non-Gaussian acyclic models (LiNGAM), it can be shown that the true underlying causal structure can be identified uniquely from merely observational data. DirectLiNGAM
  algorithm is a well-known solution to learn the true causal structure in high dimensional setting. DirectLiNGAM algorithm executes in a sequence of iterations and it performs a set of comparisons between pairs of variables in each iteration. Unfortunately, the runtime of this algorithm grows signficiantly as the number of variables increases. 
In this paper, we propose a parallel algorithm, called ParaLiNGAM, to learn casual structures based on DirectLiNGAM algorithm. We propose a threshold mechanism that can reduce the number of comparisons remarkably compared with the sequential solution. Moreover, in order to further reduce runtime, we employ a messaging mechanism between workers and derive some mathematical formulations to simplify the execution of comparisons. We also present an implelemntation of ParaLiNGAM on GPU, considering hardware constraints. Experimental results on synthetic and real data show that the implementation of proposed algorithm on GPU can outperform DirectLiNGAM by a factor up to $4600$ X.
 
\end{abstract}

}

\maketitle

\IEEEdisplaynontitleabstractindextext

%
\IEEEpeerreviewmaketitle

\section{Introduction}
\label{sec:intro}

%
%
%
%
%
%
\IEEEPARstart{D}{iscovering} the underlying causal mechanism in various natural phenomena or human social behavior is one of the primary goals in artificial intelligence and machine learning. For instance, we may be interested in recovering causal relationships between different regions of brain by processing fMRI signals \cite{huang2021diagnosis,sanchez2019estimating} or estimating causal strengths between genes in a gene regulatory networks (GRN) by observing gene expression levels \cite{marbach2012wisdom,haury2012tigress}. Having access to such causal relationships can enable us to answer to interventional or counter-factual questions which has broad impacts on designing a truly intelligent system \cite{pearl2018book}. The golden standard for the causal discovery is through conducting controlled experiments. Unfortunately, performing experiments in a system might be too costly or even infeasible \cite{ghassami2018budgeted}. As a result, there have been extensive studies in the literature of causality to recover causal relationships from merely observational data \cite{peters2017elements}.

Causal relationships among a set of variables can be represented by a directed acyclic graph (DAG) where there is a directed edge from variable $X$ to variable $Y$ if $X$ is a direct cause of $Y$. From the observational distribution, it can be shown that the true underlying causal graph can be recovered up to a Markov equivalence class (MEC) \cite{koller2009probabilistic}. There are two main approaches for recovering an MEC: constraint-based and score-based approaches. In the constraint-based approach, the MEC is identified by performing sufficient number of conditional independence (CI) tests over the observational distribution. PC \cite{spirtes2000causation} is a well-known algorithm for performing such CI tests in an efficient manner. PC algorithm runs in polynomial time to recover MEC if the maximum degree of causal graph is bounded by a constant. In the score-based approaches, the goal is to find the class of graphs maximizing a likelihood based score. Greedy equivalence search (GES) \cite{chickering2002optimal}  is one of the main score-based algorithms which reconstructs MEC by adding edges in a greedy manner.

As we mentioned above, without further assumption on the causal model, one can recover the causal graph up to an MEC. In order to uniquely recover the causal graph, we need to consider further assumptions on the causal mechanisms. For instance, if the causal mechanisms are non-linear and exogenous noises are additive, then the causal structure can be identified uniquely \cite{hoyer2008nonlinear}. Moreover, if the causal mechansims are linear, we can still recover the causal graph uniquely if the additive exogenous noises are non-Gaussian \cite{shimizu2006linear}. This assumption on the model is commonly called linear non-Gaussian acyclic model (LiNGAM) \cite{shimizu2006linear}. In ~\cite{shimizu2006linear}, an algorithm based on independent component analysis (ICA), commonly called ICA-LiNGAM, has been proposed which recovers the true causal graph under LiNGAM model. Later, a regression based method, called DirectLiNGAM ~\cite{shimizu2011directlingam}, has been presented to mitigate issues in using ICA algorithm. DirectLiNGAM algorithm has two main steps. In the first step, a causal order is obtained over the variables in the system. To do so, we compare any pair of variables like $X$ and $Y$, by regressing $Y$ on $X$ and checking whether the residual is independent of $Y$. A score is computed to measure the amount of dependency ~\cite{hyvarinen2013pairwise}. Afterwards, we select a variable that is most independent of its residuals, i.e., having minimum score among remaining variables and then append it to the causal order. We call this variable in each iteration as the root variable. Next, we remove this variable from the system by regressing it out and repeat the same procedure above until no variable is remained. After obtaining the causal order, in the second step, we perform multiple linear regressions based on the causal order in order to recover the underlying causal graph.

The executions of constraint-based or score-based algorithms might become too time-consuming as the number of variables increases in the system \cite{zarebavani2019cupc}. There have been some recent efforts to accelerate causal structure learning algorithms on multi-core machines. 
In the constraint-based approach,  in~\cite{le2015fast}, Le et al. implemented a parallel  version for a variant of PC algorithm (called PC-stable) on multi-core CPUs which reduces the runtime by an order of magnitude. Madsen et al. \cite{madsen2017parallel} proposed a method to perform conditional independence tests in parallel for PC algorithm. For the case of using GPU hardware, Schmidt et al. \cite{schmidt2018order} proposed a method to parallelize a small part of PC-stable algorithm. In \cite{zarebavani2019cupc}, Zare et al. proposed a GPU-based parallel algorithm for accelerating the whole PC-stable algorithm. The proposed algorithm parallelizes conditional independence tests over the pairs of variables or the conditional sets. Experimental results showed a significant speedup ratio up to 4000 in various real dataset. In \cite{schmidt2019out}, Schmidt et al. devised an out-of-core solution for accelerating PC-stable in order to handle extremely high-dimensional settings. Later, for discrete data, Hagedorn and Huegle \cite{hagedorn2021gpu} proposed a parallel PC algorithm on GPU for learning causal structure. Recently, Srivastava et al. \cite{srivastava2020parallel} presented a parallel framework to learn causal structures based on discovering Markov Blankets.

In score-based approach, Ramsey et al. \cite{ramsey2017million} proposed fast GES algorithm which accelerates updating score by caching scores of previous steps. They also implemented a parallel version of it on multi-core CPUs. Furthermore, there is a recent parallel solution for other search algorithms in the score based approach \cite{lee2019parallel}. 

There are some recent studies with the main focus on evaluating the performance of causal structure learning algorithms in recovering the true underlying causal graph \cite{scutari2018learns,heinze2018causal}. It has been shown that LiNGAM algorithm has comparable or better performance than most existing methods and it is more suitable for high dimensional settings \cite{heinze2018causal}. Unfortunately, the runtimes of both variants of LiNGAM algorithm (ICA-LiNGAM or DirectLiNGAM) grow significantly as the number of variables increases. Thus, the current sequential implementations cannot be utilized for dataset with large number of variables. To the best of our knowledge, there is no previous parallel implementation of LiNGAM algorithm. In this paper, we propose a parallel algorithm, which we call ParaLiNGAM, for learning causal structure based on DirectLiNGAM algorithm. Our experiments show that the first step of DirectLiNGAM is computationally intensive and we focus on accelerating this step in this paper. Similar to DirectLiNGAM, we obtain the causal order in a number of iterations sequentially while in each iteration, we parallelize the process of finding the root variable.  
The main contributions of the paper are given as follows: 

\begin{itemize}
\item 
We propose a threshold mechanism in order to reduce the number of comparisons in each iteration. In this mechanism, we consider an upper limit on the score of root variable and whenever a variable exceeds this limit, we do not perform further comparisons corresponding with that variable. Our experiments show that the threshold mechanism can save up to $93.1\%$ comparisons that we have in DirectLiNGAM.
\item When we compare variable $X$ with $Y$, a part of computation is similar to the case that we are comparing $Y$ with $X$ in the reverse direction. Thanks to this observation, we employ a messaging mechanism in order to avoid performing redundant computations which reduces runtimes by a factor of about two.
\item We derive the mathematical formulations for normalizing and also regression which are frequently utilized in DirectLiNGAM algorithm. These mathematical formulations enable us to reduce the runtime and use the memory more efficiently.   

\item We evaluate ParaLiNGAM on various synthetic and real data. Experimental results show that the proposed algorithm can reduce runtime of DirectLiNGAM significantly by a factor up to $4657$.
\end{itemize}

The rest of this paper is organized as follows. In Section \ref{sec:prelim}, we review some preliminaries on structural equal models, LiNGAM model, and DirectLiNGAM algorithm. In Section \ref{sec:alg}, we present  ParaLiNGAM algorithm for learning causal structures in LiNGAM model. We provide some implementation details in Section \ref{sec:alg2}. We evaluate the performance of ParaLiNGAM algorithm in Section \ref{sec:exp}. Finally, we conclude the paper in Section \ref{sec:conc}.

\section{Preliminaries}
\label{sec:prelim}

\subsection{Structural equation models}
\label{sec:prelim:sem}
Structural equation models (SEMs) are mathematical models that can be used to describe the data-generating process and causal relations of variables \cite{bollen1989structural, pearl2000models}. In particular, SEMs consists of a collection of $p$ equations where $p$ is the number of variables in the system. The causal mechanism of assigning values to the variable $X_j$, $1\leq j \leq p$, can be written as follows:
\begin{equation}
	X_j= f_j(PA_j, N_j),
\end{equation}
where $PA_j$ are called parents of $X_j$ and have direct cause on it. Moreover, $N_j$ is the exogenous noise corresponding to variable $X_j$. Exogenous noises are generated outside of the model and their data-generation processes are not modeled in the SEM.
We can represent causal relationships among the variables in an SEM by a directed graph where there is a directed edge from $X_i$ to $X_j$ if  $X_i \in PA_j$.
\begin{Example}
Consider the following SEM: 
\begin{figure}[tp]
	\centering
	\includegraphics[width = 0.45 \columnwidth]{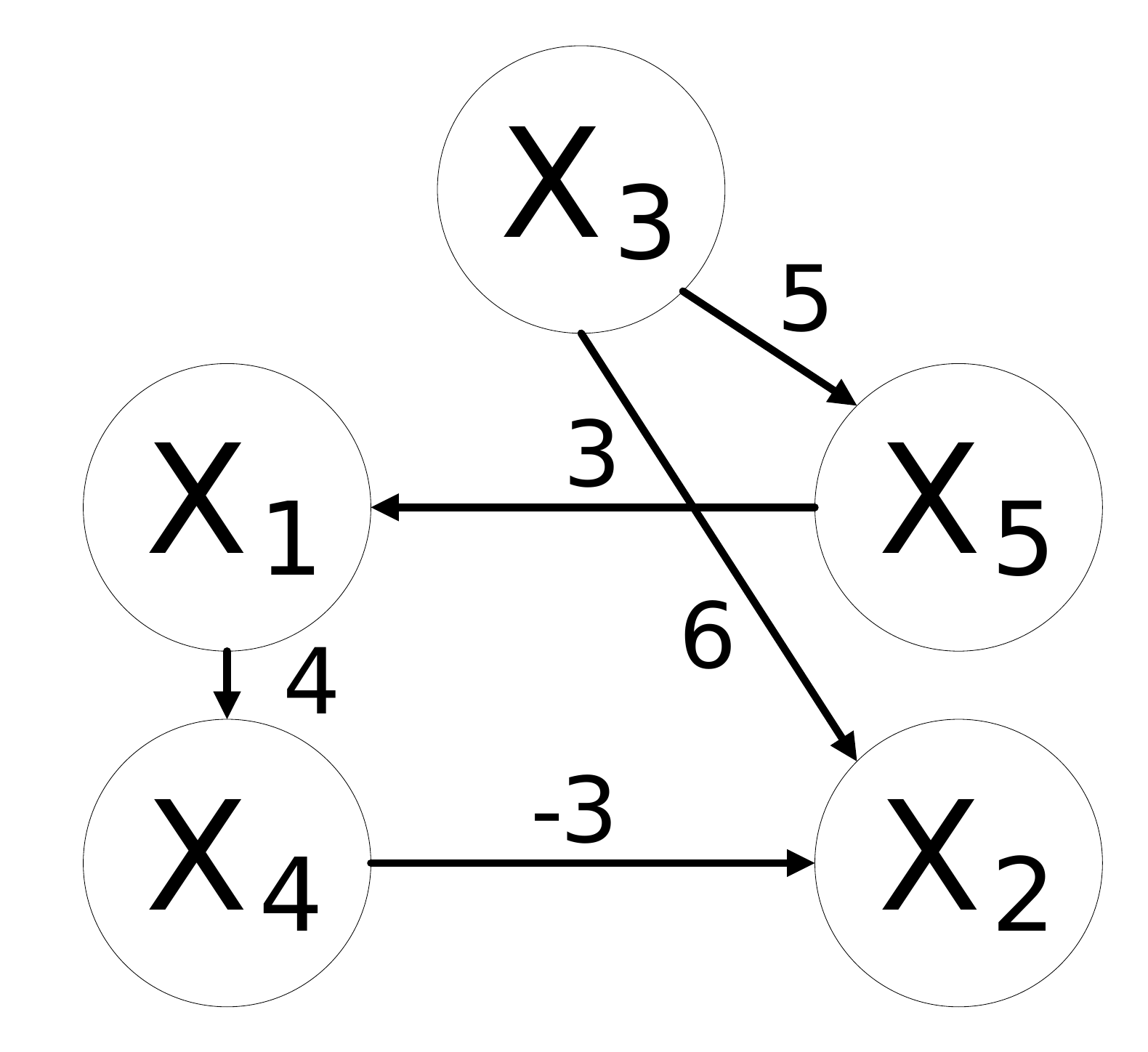}	
	\caption{Example of an SEM.}
	\label{fig:pre:SEM} 
\end{figure}
\begin{equation}
	\begin{gathered}
		X_3 = N_3, \\
		X_5 = f_5(X_3, N_5), \\
		X_1 = f_1(X_5, N_1), \\
		X_4 = f_4(X_1, N_4), \\
		X_2 = f_2(X_3, X_4, N_2),
	\end{gathered}
	\label{eq:sem}
\end{equation}
where the corresponding causal graph is illustrated in \mRefFig{fig:pre:SEM}. As can be seen, there is a directed edge from a direct cause to its effect. For instance, $X_3$ is the direct cause of $X_5$ and there is a directed edge from $X_3$ to $X_5$. 

\label{ex:base}
\end{Example}


\subsection{Linear Non-Gaussian Acyclic Model (LiNGAM)}
 One of the common assumption in the literature of causality is that the causal relations between variables are acyclic, i.e., the corresponding causal graph is a \textbf{directed acyclic graph} (\textbf{DAG})\footnote{A directed acyclic graph is a graph whose edges are all directed and there is no directed
cycle in the graph.}. By this assumption, there is always a \textbf{causal order} of variables $X_i$, $i\in\{1,\dotsc,p\}$, in the DAG so that no latter variable in the causal order has a direct path to any earlier variable. We denote the position of each variable $X_i$ in the causal order by $k(i)$. For instance, for the causal graph in \mRefFig{fig:pre:SEM}, $k=[3,5,1,4,2]$
 is a causal order.

 
As an additional assumption, one can consider that the functional relations of variables are linear. Thus, the model can be reformulated as follows:
\begin{equation}
	X_i = \sum_{k(j)<k(i)}b_{ij}X_{j} + N_i,
\label{eq:base_linear}
\end{equation}
where $b_{ij}$ is the causal strength representing magnitude of direct causation from $X_j$ to $X_i$. Furthermore, it is assumed that the exogenous noises have zero mean, non-zero variance, and are independent of each other (i.e., no latent confounder is in the system). We can rewrite Eq. \ref{eq:base_linear} in the matrix form as follows:
\begin{equation}
X = BX + N,
\label{eq:base_mat}
\end{equation}
where $X$ and $N$ are $p$-dimensional random vectors, and $B$ is a $p \times p$ matrix of causal strengths. For instance, the SEM of \mRefFig{fig:pre:SEM} can be written as follows:
\begin{equation}
\begin{bmatrix}
X_1 \\ X_2 \\ X_3 \\ X_4 \\ X_5
\end{bmatrix}
=
\begin{bmatrix}
0 & 0 & 0 & 0 & 3 \\
0 & 0 & 6 & -3 & 0 \\
0 & 0 & 0 & 0 & 0 \\
4 & 0 & 0 & 0 & 0 \\
0 & 0 & 5 & 0 & 0 \\
\end{bmatrix}
\begin{bmatrix}
X_1 \\ X_2 \\ X_3 \\ X_4 \\ X_5
\end{bmatrix}
+
\begin{bmatrix}
N_1 \\ N_2 \\ N_3 \\ N_4 \\ N_5
\end{bmatrix}
,
\label{eq:base_mat_ex}
\end{equation}
where zero entries of $B$ show the absence of directed edges. It can be shown that a simultaneous permutations of rows and columns of matrix $B$ according to a causal order can convert it to a \textbf{strictly lower triangular} matrix, due to the acyclicity assumption \cite{bollen1989structural}.
In the example in \mRefFig{fig:pre:SEM}, we can rewrite equations in the following form to make the matrix $B$ strictly lower triangular:
\begin{equation}
\begin{bmatrix}
X_3 \\ X_5 \\ X_1 \\ X_4 \\ X_2
\end{bmatrix}
=
\begin{bmatrix}
0 & 0 & 0 & 0 & 0 \\
5 & 0 & 0 & 0 & 0 \\
0 & 3 & 0 & 0 & 0 \\
0 & 0 & 4 & 0 & 0 \\
6 & 0 & 0 & -3 & 0 \\
\end{bmatrix}
\begin{bmatrix}
X_3 \\ X_5 \\ X_1 \\ X_4 \\ X_2
\end{bmatrix}
+
\begin{bmatrix}
N_3 \\ N_5 \\ N_1 \\ N_4 \\ N_2
\end{bmatrix}
.
\label{eq:base_mat_ex_prder}
\end{equation}

 It can be shown that the causal structure cannot be recovered uniquely if the distributions of exogenous noises are Gaussian ~\cite{pearl2000models}.
However, in ~\cite{shimizu2006linear}, it has been proved that the model can be fully identified from observational data if all the exogenous noises are non-Gaussian. They called the non-Gaussian version of the linear acyclic SEM,  \textbf{Linear} \textbf{Non-Gaussian} \textbf{Acyclic Model} (\textbf{LiNGAM}). In the rest of this paper, we assume the model of data generation obeys the assumptions in LiNGAM.

\subsection{Causal Structure Learning Algorithms for LiNGAM}
\label{sec:baseLiNGAM}
ICA-LiNGAM ~\cite{shimizu2006linear} was the first algorithm for the LiNGAM model, which applies an independent component analysis (ICA) algorithm to observed data and try to find the best strictly lower triangular matrix for $B$ that fits the observed data. This algorithm is fast due to well-developed ICA techniques. However, the algorithm has several drawbacks, e.g., getting stuck in local optimal, scale-dependent calculations, and usually estimating dense graph even for sparse ground truth causal graphs~\cite{shimizu2011directlingam}.

\textbf{DirectLiNGAM} was proposed in~\cite{shimizu2011directlingam}, in order to resolve ICA-LiNGAM's issues and converges to an acceptable approximation of matrix $B$ in a fixed number of steps. 
Also, we can provide prior knowledge to the algorithm, which can improve the performance of recovering the correct model. However, computation cost of DirectLiNGAM is more than ICA-LiNGAM, and it cannot be applied on large graphs\cite{shimizu2011directlingam}.

DirectLiNGAM algorithm consists of two main steps: In the first step, the causal order of variables is estimated by repeatedly searching for a root in the remaining graph and regressing out its effect on other variables. In the second step, causal strengths are estimated by using some conventional covariance-based regression according to the recovered causal order~\cite{shimizu2011directlingam}. Experimental results show that the second step is fairly fast since we are only performing linear regressions. However, the first step is computationally intensive and we focus on accelerating this part in this paper. 


\begin{algorithm}[tp]
	\begin{algorithmic}[1]
		\Input $\mathcal{X}$
		\Output $K$
		\State $U =\{1,\cdots, p\}$ 
		\State $K=\emptyset$
		\Repeat
			\State $root = FindRoot(\mathcal{X}, U)$
			\State Append $root$ to $K$
			\State Remove $root$ from $U$
			\State $\mathcal{X} = RegressRoot(\mathcal{X}, U, root)$
		\Until{$U$ is not empty}
		\State Estimate causal strengths $B$ from $K$
	\end{algorithmic}
	\caption{DirectLiNGAM} 
	\label{alg:baseLiNGAM}
\end{algorithm}

The description of DirectLiNGAM algorithm is given in \mRefAlg{alg:baseLiNGAM}. The input of the algorithm is matrix $[\mathcal{X}]_{p\times n}$
whose $i$-th row, $x_i$, contains $n$ samples from variable $X_i$.
The output of algorithm is $K$, a causal order list of variables. First, we initialize $U$ by a list of all variables' indecies and set $K$ to an empty list (lines $1-2$). Next, $K$ is going to be filled with variables from $U$ using a comparison between variables to form a causal order. Recovering a causal order takes $p$ (size of $U$) iterations. In each iteration, the most independent variable in $U$ ($root$ of that iteration) is determined by the $FindRoot$ function. Then the $root$ moves from $U$ to $K$. Next, the data of remaining variables in $U$ are updated by regressing them on the root ($RegressRoot$ function), which has been shown that it preserves correct causal orders in the remaining part ~\cite{shimizu2011directlingam}. Finally, the matrix B is recovered from the causal order in $K$s using conventional covariance-based regression methods.



\begin{algorithm}[tp]
	\begin{algorithmic}[1]
		\Input $\mathcal{X}, U$
		\Output $root$
		\If {$U$ has only one element}
			\State return the element
		\EndIf
		\State $\mathcal{S} = [\textbf{0}]_{|U|}$ 
		\For {$i$ in $U$}
			\For {$j$ in $U\backslash\{i\}$}
				\State $Normalize(x_i)$
				\State $Normalize(x_j)$
 			
				\State $r_i^{(j)} =  Regress(x_i, x_j)$
				\State $r_j^{(i)} =  Regress(x_j, x_i)$
				\State $Normalize(r_i^{(j)})$
				\State $Normalize(r_j^{(i)})$
				\State $\mathcal{S}[i] += \min\{0, \mTest(x_i, x_j, r_i^{(j)}, r_j^{(i)})\}^2$
			\EndFor
		\EndFor
		\State $root=$ $U[arg\min(Scores)]$
	\end{algorithmic}
	\caption{$FindRoot$} 
	\label{alg:baseLiNGAM_FR}
\end{algorithm}

The purpose of $FindRoot$ function (see \mRefAlg{alg:baseLiNGAM_FR}) is to find most independent variable from its residuals by comparing all pairs of variables given in $U$. Each variable in $U$ has a score with initial value of zero and all of them are stored in an array called $\mathcal{S}$ (line $4$).
First, samples of each variable are normalized. Next, each variable $X_i$ is regressed on any other variable like $X_j$ in $U$. Afterwards, the regressed values are normalized. Finally, an independence test $\mTest$ is performed and its result is added to a score of variable $X_i$. In ~\cite{hyvarinen2013pairwise}, a likelihood ratio test is proposed that assigns a real number to the pair of variables:
\begin{equation}
\mTest(x_i, x_j, r_i^{(j)}, r_j^{(i)}) = H(x_j) + H(r_i^{(j)}) - H(x_i) - H(r_j^{(i)}),
\label{eq:Itest}
\end{equation}
where $H$ is differential entropy, which can be approximated by computationally simple function as follows ~\cite{hyvarinen2013pairwise, hyvarinen1998analysis}:

\begin{equation}
\!\begin{multlined}[t]
\hat{H}(u) = H(v) - k_1[E\{\log\cosh(u)\}- \beta]^2 \\
 - k_2[E\{u\exp(-u^2/2)\}]^2.
\end{multlined}
\label{eq:HApp}
\end{equation}

In above equation, $H(v) = \frac{1}{2}(1 + \log2\pi)$ is the entropy of the standardized Gaussian distribution, and the other constants can be set to:
\begin{center}
$k_1 \approx 79.047,$\\
$k_2 \approx 7.4129,$\\
$\beta \approx 0.37457.$\\
\end{center}

A positive/negative value of $\mTest$ indicates the independence/dependence of that variable compared to the other one. For aggregating score for each variable and determining total independence of a variable from others, only the amount of dependence is considered. 
In other words, in this method, only negative value of $I$ are considered and its square is added to the score.
The variable with minimum score is selected as the root variable. 
Please note that we consider lines $9-13$ as \textbf{$Compare$} function and use this as the based function for comparing two variables in next sections.



%

\section{ParaLiNGAM}
\label{sec:alg}

In this section, we present the ParaLiNGAM algorithm for accelerating computationally-intensive part of DirectLiNGAM without changing its accuracy.

\begin{figure*}[tp]
	\centering
	\includegraphics[width = \textwidth]{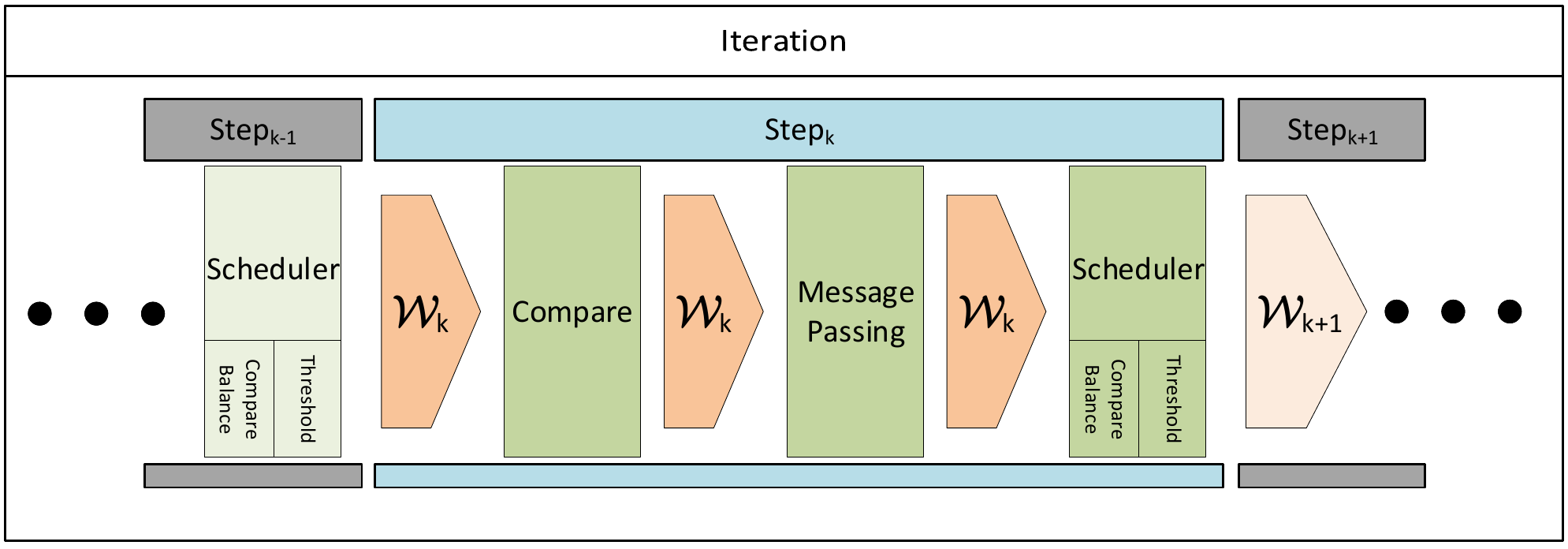}	
	\caption{Procedure of one iteration in ParaLiNGAM: Each iteration is divided into steps.
	Each step has three parts: Compare, Message Passing, Scheduler, which are accomplished with pre-selected workers from the previous step ($\mathcal{W}_k$). Details are discussed in Section \ref{sec:alg}.}
	\label{fig:alg:base}
\end{figure*}

As mentioned before, DirectLiNGAM discovers causal order in $p$ iterations. Moreover, these iterations are consecutive, i.e., an iteration cannot be started unless the previous has been already finished. Herein, ParaLiNGAM is also executed in $p$ iterations. \mRefFig{fig:alg:base} illustrates the procedure of one iteration.

In each iteration, computations of each variable are assigned to a specific \textbf{worker}. Iterations are broken into \textbf{steps}. In each step $k$, a subset of all workers($\mathcal{W}$) in an iteration, which is denoted by $\mathcal{W}_k$, is selected to start comparing themselves with other workers (Compare part). Next, the workers inform each other about their computations by sending messages (Message Passing part).
Finally, the \textbf{scheduler} gathers all workers' scores and selects some of the workers for the next step, i.e., $\mathcal{W}_{k+1}$. 

\begin{algorithm}
	\begin{algorithmic}[1]
		\Input $\mathcal{X}$
		\Output $K$
		\State $K=\emptyset$
		\State $U =[1,\cdots, p]$
		\State Par: $NormalizeData(\mathcal{X})$
		\State Par: $ \Sigma = CalculateCovMat(\mathcal{X})$
		\Repeat
			\State Par: $root = ParaFindRoot(\mathcal{X}, U, \Sigma)$
			\State Append $root$ to $K$
			\State Remove $root$ from $U$
			\State Par: $\mathcal{X} = UpdataData(\mathcal{X}, U, \Sigma, root)\hfill//$Sec\ref{sec:math}
			\label{line:RegressRoot}
			\State Par: $\Sigma = UpdateCovMat(\mathcal{X}, U, \Sigma, root) \hfill//$ Sec\ref{sec:math}
			\label{line:UpdateCovMat}
		\Until{$U$ is not empty}
	\end{algorithmic}
	\caption{ParaLiNGAM} 
	\label{alg:ParaLiNGAM}
\end{algorithm}

The description of ParaLiNGAM is given in \mRefAlg{alg:ParaLiNGAM}. The lines starting with ``Par" show that they are executed in parallel. ParaLiNGAM algorithm is executed similar to DirectLiNGAM. First, we initialize $U$ and $K$ (lines $1-2$). Next, we find a casual order in lines $5-11$. General procedure of this algorithm is same as DirectLiNGAM, however, computations' details have been changed which we discuss them in the sequel.

In DirectLiNGAM, samples of all variables have to be normalized in $FindRoot$ function (lines $7-8$ in \mRefAlg{alg:baseLiNGAM_FR}). For sake of efficiency, all variables are normalized simultaneously in line $3$ of \mRefAlg{alg:ParaLiNGAM} for the first iteration. For the next iterations, this task is done with $UpdateData$ function (line $9$ in \mRefAlg{alg:ParaLiNGAM}).

Regressing variables on each other, is a frequent task in DirectLiNGAM which is performed in $Compare$ (lines $9-10$ in \mRefAlg{alg:baseLiNGAM_FR}) and $RegressRoot$ (line $7$ in \mRefAlg{alg:baseLiNGAM}) functions, and it needs variables' covariance matrix. Hence, it is desirable to store covariance matrix (which we denote it by $\Sigma$) in each iteration to avoid redundant computations. In \mRefAlg{alg:ParaLiNGAM}, the first covariance matrix of variables is calculated in line $4$, and it is updated in each iteration (line $10$). 
Furthermore, we will show in Section \ref{sec:math} how to reuse computations from previous iterations in normalizing data and obtaining the covariance matrix which results in reducing the computational complexity without degrading the accuracy.
 

We discussed the summary of changes in \mRefAlg{alg:baseLiNGAM}. Now, we are ready to explain these changes in more details. First, we present a parallel solution for finding root in each iteration ($ParaFindRoot$ function in Algorithm \ref{alg:ParaLiNGAM_FR}).

\begin{algorithm}
	\begin{algorithmic}[1]
		\Input $\mathcal{X}, U, \Sigma$
		\Output $root$
		\item[\textbf{\# of Workers:}] $|U|$
		\If {$U$ has only one element}
			\State return the element
		\EndIf
		\State $r = |U|$
		\State $\mathcal{W} = [1, 2, 3, ..., r]$
		\State $\mathcal{S} = [\textbf{0}]_{r}$
		\State $\mathcal{M} = [\emptyset]_{r\times r}$
		\State $\mathcal{D} = diag([True]_r)$
		\State $State = \{U, \Sigma, r, \mathcal{S}, \mathcal{M}, \mathcal{D}, \gamma \}$
		\State $\mathcal{W}' = \mathcal{W}$
		\State $\mathcal{C} = \textbf{1}_{r}$
		\Repeat
		
				\State Par: $\mathcal{S}[w] += Compare(w, \mathcal{C}[w], State)$
				\State Par: $\mathcal{S}[w] += CheckMessages(w, State)\hfill//$Sec\ref{sec:messaging}
			\State $finished, \mathcal{W'}, \mathcal{C} = Scheduler (\mathcal{W},  \mathcal{C}, State)\hfill//$Sec\ref{sec:scheduler}
		\Until{$finished == True$}
		\State $root=U[\arg\min(\mathcal{S})]$
	\end{algorithmic}
	\caption{$ParaFindRoot$} 
	\label{alg:ParaLiNGAM_FR}
\end{algorithm}

In order to find root in DirectLiNGAM (\mRefAlg{alg:baseLiNGAM_FR}), all variables compare themselves with other ones (line $5$). In the parallel version, we assign each variable to a worker to perform its computations. In other words, instead of a $for$ statement in line $5$ of \mRefAlg{alg:baseLiNGAM_FR}, we have workers, which can work in parallel to perform the comparisons. Moreover, comparing each variable/worker to other variables/workers is divided into steps instead of iterating on all variables (line $6$ \mRefAlg{alg:baseLiNGAM_FR}). Next, we discuss the details of presented solution for finding root (\mRefAlg{alg:ParaLiNGAM_FR}).

The description of $ParaFindRoot$ function is given in  \mRefAlg{alg:ParaLiNGAM_FR}. First, we define $r$ as the number of remaining variables in this iteration, which equals to the size of $U$ (line $4$). In each iteration, every pair of variables have to be compared to find the root. Moreover, these comparing procedure is independent of each other. We can use $r$ workers, and worker $i$ is responsible for performing variable $U[i]$'s computations.
For simplicity of notations, from now on, we denote workers by the corresponding variables assigned to them. We define $\mathcal{W}$ as a list of all workers' indices in an iteration (line $5$). In line $6$,  $\mathcal{S}$ is initialized the same as the main algorithm (line $4$ in \mRefAlg{alg:baseLiNGAM_FR}) to store scores. Each worker might have useful information for the other workers, which can be shared with a \textbf{messaging} mechanism. To do so, we define $\mathcal{M}$ which is $r\times r$ matrix filled with $\emptyset$ (line $7$). Worker $i$ can send a message to worker $j$ by writing in $\mathcal{M}[j][i]$. More details of the messaging mechanism and its effect on the performance of algorithm will be discussed in \mRefSec{sec:messaging}. Note that matrix $\mathcal{M}$ is just temporary memory for messaging and it resets in each step. 
Hence, another variable is required for evaluating the progress of an iteration.
For this purpose, in line $8$, we define a $r \times r$ matrix $\mathcal{D}$ in which diagonal entries are initially $True$ while others are $False$, to monitor workers' progress. Worker $i$  writes $True$ in $\mathcal{D}[i][j]$ after comparing itself with worker $j$.
For sake of brevity, we collect all variables defined in Algorithm \ref{alg:ParaLiNGAM_FR}, and the threshold $\gamma$ with a small value (Section \ref{sec:thresh}, \ref{sec:scheduler}) in set $State$ (line $9$).

As mentioned before, finding root in each iteration is divided into steps. In each step, some workers are selected, and each of them has to compare itself to another worker. Selected workers are indicated with $\mathcal{W}'$ (line $10$). In the first step of each iteration, all of workers have same priority. Hence, $\mathcal{W}'$ equals $\mathcal{W}$. Moreover, list of target workers to be compared with is defined as $\mathcal{C}$ (line $11$), and worker $i$ compares itself with $\mathcal{C}[i]$. At the beginning of each iteration, $\mathcal{C}$ is initiated with a list filled with $1$, which means all of the workers start comparing themselves with the first worker.

In each step, first, selected workers compare themselves with the assigned workers and send a message (line $13$). Comparing and sending a message is an independent task for each worker and can be performed in parallel. Then, workers check for new messages from others and update their scores (line $14$). Afterwards, $Scheduler$ selects workers for the next step according to this step's status. Moreover, it modifies $\mathcal{C}$ for selected workers (line $15$) and determines whether to terminate an iteration after checking its $State$. Finally, similar to \mRefAlg{alg:baseLiNGAM_FR}, the worker with minimum score is chosen as the root of iteration (line $17$). 

There are still some implementation details that will be discussed in the next parts. More specifically, 
messaging between workers will be discussed in \mRefSec{sec:messaging}. The details of \textbf{threshold} mechanism which determines $\mathcal{W}'$, the set of selected workers, is given in \mRefSec{sec:thresh}. The scheduling of workers based on the threshold mechanism is discussed in \mRefSec{sec:scheduler}. Mathematical simplification for accelerating $UpdateData$ and $UpdateCovMat$ process is also discussed in \mRefSec{sec:math}. Finally, implementation of this algorithm on GPU and further details on some parts of solution are given in \mRefSec{sec:alg2}.

\subsection{Messaging}
\label{sec:messaging}
In DirectLiNGAM, as mentioned in \mRefSec{sec:baseLiNGAM}, computation of test $\mTest$ for $x_i \rightarrow x_j$ has some similarities to the ones for $x_i \leftarrow x_j$. It is worth mentioning, this property is not just for $\mTest$, and this technique can be used for some other tests \cite{hyvarinen2013pairwise}.
When worker $i$ compares itself to worker $j$ (computing $\mTest(x_i, x_j, r_i^{(j)}, r_j^{(i)})$), it can also compute the test in the reverse direction ($\mTest(x_j, x_i, r_j^{(i)}, r_i^{(j)})$), which is worker $j$'s task. Hence, we can assign full comparison to just one worker, and after finishing each comparison, the worker who performed the test, has to inform the other worker about its result by sending a message. With messaging, which does not add computation load
we can halve the comparisons (from $p(p-1)$ to $p(p-1)/2$). To use this mechanism, every active worker in a step has to send a message after finishing its comparison (line 13 in \mRefAlg{alg:ParaLiNGAM_FR}), and each worker checks for their messages (which can be from any $r-1$ other workers in the whole iteration) with $CheckMessages$ function.

\begin{algorithm}
	\begin{algorithmic}[1]
		\Input $w, State$
		\Output $score$
		\State $score = 0$
		\For{$~~(i=1;~~ i <= r;~~ i+=1)$}
			\If{$\mathcal{M}[w,i] != \emptyset$}
				\State $score += \mathcal{M}[w, i]$
				\State$\mathcal{D}[w, i] = True$
				\State $\mathcal{M}[w, i] = \emptyset$
			\EndIf
		\EndFor
			
	\end{algorithmic}
	\caption{$CheckMessages$}
	\label{alg:CheckMessages}
\end{algorithm}

The description of $CheckMessages$ function is given in \mRefAlg{alg:CheckMessages}. First, we define $score$ as variable for sum of scores. In lines 2-3, worker $w$ checks for new messages. If another worker, say $i$, has sent a message, first, $score$ is updated (line 4). Next, worker $w$ marks the sender worker ($i$) as "done" by writing $True$ in $\mathcal{D}[w, i]$ (line 5). Finally, message is replaced with $\emptyset$ to prevent it from recalculation in next steps. The senders and receivers might not be active simultaneously in one step. As a result, workers consider messages from whole workers in an iteration and not just from active workers in the current step.

\subsection{Threshold}
\label{sec:thresh}
As mentioned earlier, we need to perform $p(p-1)/2$ comparisons in each iteration. However, all of them are not necessary. Suppose that the final score of the root in an iteration is $0.05$. In this case, we can terminate any worker's computation whose score has reached $0.05$.
To reduce the number of comparisons, we consider an upper bound for the score which we call it threshold, and assume that root's score will probably be less than this threshold. If that is the case, the iteration is over and we can choose the root when at least one worker could finish its comparisons without reaching the threshold, while all other workers reach it without completing their tasks. 

Herein, the main issue is how to choose a proper threshold.
To overcome this issue, first, we choose a small value for the threshold. Furthermore, we terminate workers who have already reached this value. Then, if all workers terminate and neither of them could finish their comparisons, we increase the threshold. We continue this procedure till iteration termination conditions satisfy, i.e., at least one worker finish all their comparisons without reaching the threshold. Consequently, using the threshold mechanism can result in reducing the number of comparisons.
%

Now, we discuss the correctness of the threshold mechanism in each iteration.
At the end of each iteration, each worker is in one of these two groups:
1) Worker's score is more than the threshold, and it may not even finish its comparisons.
2) Worker finishes its comparisons, and still, its score is below the threshold.
Workers in the first group have higher scores than the threshold even they continue their comparisons. The root is a worker with a minimum score. As a result, it is always chosen from the second group of workers. In fact, we select the root from workers in the second group which has the minimum score among them. Hence, early termination of the first group of workers does not affect algorithm results.
Details of reducing number of comparisons by the threshold mechanism are discussed in \mRefSec{sec:scheduler}.

\subsection{Scheduler}
\label{sec:scheduler}

In this part, we explain how to schedule workers with the threshold mechanism. At the end of each step, the scheduler has to decide whether or not to finish current iteration.

\begin{algorithm}
	\begin{algorithmic}[1]
		\Input $\mathcal{W}, \mathcal{C}, State$
		\Output $finish, \mathcal{W'}, \mathcal{C}$
		\State $finish = False$
		\ForAll {$ w \in \mathcal{W}$} // checking the termination 
			\If{$\mathcal{S}[w] < \gamma$}
					\If {$\mathcal{D}[w,:]$ has at least one $False$}
						\State $finish = False$
						\State \textbf{Break}
					\Else
						\State $finish = True$
					\EndIf
			\EndIf	
		\EndFor
		\If {$finish$}
			\State \textbf{return}
		\EndIf
		\While{$\mathcal{S}[w] > \gamma, ~~ \forall w \in \mathcal{W}$} 
			\State $\gamma = \gamma \times c$ // updating threshold
		\EndWhile
		\State $\mathcal{W'} = [ w \in \mathcal{W}~~|~~ \mathcal{S}[w] < \gamma]$
		\ForAll {$ w \in \mathcal{W'}$}
			\Repeat
				\State $\mathcal{C}[w] += 1$
			\Until{$(\mathcal{D}[w,\mathcal{C}[w]] == False ~~ \& \& ~~ \mathcal{C}[\mathcal{C}[w]]!=w) ~~ | | ~~ \mathcal{C}[w] > |\mathcal{W}| $}
		\EndFor
		\State $\mathcal{W'} = \mathcal{W'}\backslash[ w \in \mathcal{W'}~~|~~ \mathcal{C}[w] > |\mathcal{W}|]$	
		
	\end{algorithmic}
	\caption{$Scheduler$}
	\label{alg:Scheduler}
\end{algorithm}

As mentioned earlier in \mRefSec{sec:thresh}, if at least one worker has finished its comparisons and its score is below than the threshold, we can terminate the iteration.

The description of scheduler is given in \mRefAlg{alg:Scheduler}. The termination of an iteration is checked in lines $1-14$. First, we define $finish$ as a flag for termination conditions (line $1$). Then, we have to find workers with scores less than the threshold (line $2-3$) and check whether they have finished their comparisons or not (line 4). If at least one worker has unfinished comparisons and its score is below than the threshold, we need to continue another step (lines $5-6$). If some workers have finished their comparisons, $finish$ is changed to $True$, and we wait for other workers' status (line $7-9$). Then, we check $finish$ value and decide whether or not to terminate the iteration (lines $12-14$).

Now we discuss the scheduler's task when an iteration is not terminated and it should be continued for another step. For the new step, we have to change threshold if it is needed (lines $15-17$). There are multiple ways to increase the value of threshold. Here, we just multiply by some constant $c$ (see \mRefSec{sec:schGPU} for more details on selecting the desirable constant $c$). We continue updating threshold until at least one of workers' scores is below than the threshold. Then, the scheduler chooses workers ($\mathcal{W}'$) and comparison targets ($\mathcal{C}$) for the next step.

Workers with scores less than the threshold are considered as the workers of new step (line $18$). $\mathcal{C}$ is updated for the selected workers in lines $19-23$. As mentioned, $\mathcal{C}$ is initialized with \textbf{$1$} and workers need to compare themselves with worker $1$ for their first step. For the next steps, we start increasing $\mathcal{C}[w]$ for worker $w$ till we find a pair ($w$, $\mathcal{C}[w]$) which their comparison has not performed yet.
In a step, it might occur that two workers compare with each other  simultaneously, which is not desirable as they are performing redundant tests. In order to prevent these cases, the scheduler also checks for repetitive pairs of comparison (second condition in line $22$).

A worker might finish its comparisons while its score was greater than the threshold, but it becomes smaller after updating threshold. Such workers have to wait for the next step. In this case, first, $\mathcal{C}[w]$ is set to a value greater than $|\mathcal{C}|$, then the scheduler omits such workers from $\mathcal{W}'$ (line $24$).

Some details of scheduling depend on implementation considerations and will be discussed in \mRefSec{sec:schGPU}.

\subsection{Mathematic Simplification}
\label{sec:math}

As mentioned earlier, DirectLiNGAM algorithm always works with normalized data. Furthermore, normalization and regression tasks are frequently performed in the algorithm. These tasks depend on computing the covariances of variables, and it would be desirable to obtain them in an efficient manner. In this section, we demonstrate that if the relationship between variables is linear, which is one of the main assumption in LiNGAM, we can estimate variance of residual of regressions and use them in normalization step (lines $7 - 8$ in \mRefAlg{alg:baseLiNGAM_FR}). Furthermore, we can estimate coefficients used in regressing variables in data updating procedure (line $7$ in \mRefAlg{alg:baseLiNGAM}).

First, we calculate  the adjusted sample variance $s^2$ for residual of a regression. Residual of $x_i$ regressed on $x_j$ (denoted by $r_i^{(j)}$) is defined as:
\begin{equation}
r_i^{(j)} = x_i - \frac{cov(x_i, x_j)}{var(x_j)} x_j, i \neq j.
\end{equation}
Moreover, if we assume that samples are normalized, it can be shown that $E[r_i^{(j)}] = 0$. Furthermore, calculating residuals only needs the covariance matrix. Consider  $b = cov(x_i, x_j)$ and assume that both variables are normalized. We can write:
\begin{equation}
	\begin{split}
	s^2 &= \frac{1}{n-1} \sum (r_i^{(j)} - E[r_i^{(j)}])^2 \\
	&= \frac{1}{n-1} \sum (x_i - b x_j)^2 = \frac{1}{n-1} \sum (x_i^2 - b^2x_j^2 - 2b x_i x_j) \\
	&= \frac{1}{n-1} (\sum x_i^2 + b^2 \sum x_j^2 - 2b\sum x_i x_j) \\
	&= var(x_i) + b^2 var(x_j) - 2b cov(x_i, x_j)\\
	&= 1 + b^2 - 2b^2 = 1 - b^2 = 1 - cov^2(x_i, x_j).
	\end{split}
	\label{eq:var_update}
\end{equation}

Thus, we showed that the variance of $r_i^{(j)}$ is equal to $1 - cov^2(x_i, x_j)$. Therefore, in order to normalize it, we just have to divide all samples to $\sqrt{1 - cov^2(x_i, x_j)}$. The details of $UpdateData$ function, which is called in \mRefAlg{alg:ParaLiNGAM}, is given in \mRefAlg{alg:UpdateData}.
\begin{algorithm}
	\begin{algorithmic}[1]
		\Input $\mathcal{X}, U, \Sigma, root$
		\Output $\mathcal{X}$
		\item[\textbf{\# of Workers:}] $|U|$
		\State $\mathcal{W} = [1, 2, 3, ..., |U|]$
		
%
%
		\State Par: \textbf{for} {$~~(i=0;~~ i < |\mathcal{X}[w,:]|;~~ i+=1)$} \textbf{do}
			\State $~~~~\mathcal{X}[w, i] = \dfrac{\mathcal{X}[w, i] - \Sigma[w, root] \mathcal{X}[root, i]}{\sqrt{1 - \Sigma^2[w, root]}}$
		\State \textbf{end for}
	\end{algorithmic}
	\caption{$UpdateData$} 
	\label{alg:UpdateData}
\end{algorithm}

Next, we calculate the covariance between two residuals, which can be used in computing the covariance matrix for the next iteration.
Suppose we have $b_1 = cov(x_i, x_{root})$ and $b_2 = cov(x_j, x_{root})$. We have:
\begin{equation}
	\begin{split}
	cov(r_i^{root}, r_j^{root}) &= \frac{1}{n-1} \sum (x_i - b_1x_{root})(x_j - b_2x_{root}) \\
	&= \!\begin{multlined}[t]
		\frac{1}{n-1} \sum (x_ix_j - b_2x_ix_{root}  - b_1 x_j x_{root} \\+ b_1b_2x_{root}^2)
	\end{multlined}\\
	&= \!\begin{multlined}[t]
		cov(x_i, x_j) - b_2 cov(x_i, x_{root}) \\- b_1 cov(x_j, x_{root}) + b_1b_2 var(x_{root})
	\end{multlined}\\
	&= cov(x_i, x_j) - b_1b_2.
	\end{split}
	\label{eq:CovUpdate}
\end{equation}
From Equations \ref{eq:CovUpdate} and \ref{eq:var_update}, we can simply update the covariance matrix in each iteration just from the covariance matrix in the previous iteration, without using variables' samples (excluding the first iteration).
Details of $UpdateCovMat$ Function, which is called in \mRefAlg{alg:ParaLiNGAM}, is given in \mRefAlg{alg:UpdateCovMat}. Please note that $r_i^{root}$ and $r_j^{root}$ are not normalized in \mRefEq{eq:CovUpdate}. Therefore, it is needed to divided the expression for the covariance by the variances of  $r_i^{root}$ and $r_j^{root}$ (line 3 in \mRefAlg{alg:UpdateCovMat}).

\begin{algorithm}
	\begin{algorithmic}[1]
		\Input $\mathcal{X}, U, \Sigma, root$
		\Output $\Sigma$
		\item[\textbf{\# of Workers:}] $|U|$
		\State $\mathcal{W} = [1, 2, 3, ..., |U|]$
		
%
		\State Par: \textbf{for} {$~~(i=0;~~ i < |\Sigma[w,:]|;~~ i+=1)$} \textbf{do}
			\State $~~~~\Sigma[w, j] = \dfrac{\Sigma[w, j] - \Sigma[w, root] \Sigma[j, root]} {\sqrt{1 - \Sigma^2[w, root]}\sqrt{1 - \Sigma^2[j, root]}}$
		\State \textbf{end for}
	\end{algorithmic}
	\caption{$UpdateCovMat$} 
	\label{alg:UpdateCovMat}
\end{algorithm}

\section{Implementation Details}
\label{sec:alg2}

Further details on the proposed parallel algorithm is presented in this section. First, a short background on CUDA is presented in Section \ref{sec:CUDA}. Further details on our GPU implementation are discussed in Sections \ref{sec:GPUImp} and \ref{sec:schGPU}.

\subsection{CUDA}
\label{sec:CUDA}

CUDA is a parallel programming API for Nvidia GPUs. GPU is a massively parallel processor with hundreds to thousands of cores.
CUDA follows a hierarchical programming model. At the top level, computationally intensive functions are specified by the programmer as CUDA \textbf{kernels}. For briefness, from now on, we use kernel instead of CUDA kernel. A kernel is specified as a sequential function for a single \textbf{thread}. The kernel is then launched for parallel execution on the GPU by specifying the number of concurrent threads. 

Threads are grouped into \textbf{blocks}. A GPU kernel consists of a number of blocks, and every block consists of a number of threads.
%

In order to identify blocks within a kernel, and threads within a block, a set of indices are used in the CUDA API, for instance, $blockIdx.x$ denotes  the block index in dimension $x$ within a kernel. 



\subsection{GPU Implementation}
\label{sec:GPUImp}
In this section, we present further details on the implementation of ParaLiNGAM on GPU hardware. 
In the ParaLiNGAM algorithm, we used workers to handle variables' tasks. In CUDA, we assign workers to blocks. Moreover, each block can divide its computations among parallel threads to improve the performance, e.g., for performing tasks like $CheckMessages$ and $Compare$. Herein, for brevity of notation, we denote $blockIdx.x$ by $w$.
Now we discuss two approaches to implement ParaLiNGAM on GPU.

The first approach is to assign fewer threads to each block and decrease the computation power of each block, in the meantime, run all of the blocks together. 
Thereby, a whole iteration can be launched by one kernel. Moreover, we can perform scheduling in GPU by considering one block for the scheduler, i.e., all blocks but one skip scheduling part based on their block IDs.

However, this solution has some drawbacks.
First, each worker requires its own exclusive memory. Hence, all workers may not fit in GPU memory. As a result, it is not scalable, and we cannot utilize it for large number of variables.
Moreover, blocks are slower due to the smaller number of threads. As a result, this approach would be time-consuming, even if all the blocks fit in the GPU.

The second approach is to run $Compare$ and $CheckMessages$ tasks in separate kernels on GPU while performing the scheduling task on the host (CPU). This approach is scalable and does not have the above problems. However, launching kernels is time-consuming. Moreover, we launch two kernels separately for each step, which is not efficient.
Unlike the first approach, we can consider some modifications to the second approach in order to improve its performance, which is discussed in the next part.

\subsection{Scheduling on GPU}
\label{sec:schGPU}

In order to resolve second approach's issues, one solution is to relax the synchronization between $Compare$ and $CheckMessages$. In Algorithm \ref{alg:ParaLiNGAM_FR}, we consider a barrier between these two functions (lines $9$ and $10$) in order to update the score of workers faster by checking their messages only after all messages are received. 
However, removing the barrier causes some workers to receive their messages later in the further steps. Hence, relaxing the synchronization causes a delay in delivery of some messages, however, it halves the kernel launching delays, which is quite beneficial. As a result, the workers (blocks) can compare and check for messages independently, and thus, we can merge $Compare$ and $CheckMessages$ kernels. Still, assigning each step to a kernel and evaluating iteration status by the host is not an efficient option and causes too many kernel calls. Therefore, the procedure of scheduling must be revised.

In ParaLiNGAM, a worker can perform just one comparison (if it is active) in each step. This limitation was due to the synchronization of $Compare$ and $CheckMessages$, which is now relaxed. 
Hence, now we can divide an iteration into steps by threshold updates, instead of performing one comparison per worker. In other words, workers now can continue their comparisons to reach the threshold instead of just performing one comparison. 
To implement this solution, we modify some of the scheduler's tasks in order to move them to the worker.  Scheduler has three tasks: checking termination of an iteration, updating the threshold, and updating $\mathcal{C}$. Now, we discuss modifications that are needed in these three tasks.

In the first task, i.e., checking iteration's termination, worker $w$ can keep track of its finished comparisons by checking $\mathcal{D}[w]$. 
Therefore, workers can continue to compare themselves with each other till they reach the threshold or finish their comparisons. Moreover, they can announce their state by $finish$ flag.

The second task is updating the threshold, which is a mechanism to divide iteration into steps.
In Algorithm \ref{alg:ParaLiNGAM_FR}, workers synchronize before updating the threshold.
As mentioned before, synchronizing a kernel is not efficient. Hence, we update the threshold outside of the kernel in the host.
In Section \ref{sec:scheduler}, constant $c$ is introduced to control the amount of change in threshold in each update. 
In particular, higher values for $c$ would cause increasing the number of comparisons. However, it can be more efficient due to fewer kernel calls. As a result, this parameter must be adjusted according to the number of tests, test duration, and launching kernel delays.

In the last task, workers can also determine their comparisons' target by checking their unfinished workers in $\mathcal{D}$.

Now we discuss changes in the main algorithms to implement them on GPU which are given in Algorithm \ref{alg:SchedulerV2} and Algorithm \ref{alg:GPUKernel}.

Main loop of Algorithm \ref{alg:ParaLiNGAM_FR}  (lines $14-18$) is replaced to Algorithm \ref{alg:SchedulerV2}. In this algorithm, $\mathcal{C}$ is modified for more efficiency, and also functions of $Compare$, $CheckMessages$ and some parts of $Scheduler$ are moved to $GPUKernel$ (Algorithm \ref{alg:GPUKernel}).

\begin{algorithm}
	\begin{algorithmic}
		\State $\mathcal{C} = [2, 3, ..., r, 1]$
		\State $finish = False$
		\Repeat
			\State GPU: $finish, State = GPUKernel(\mathcal{W},
			State)$
			\State $\gamma *= c$
		\Until{$finish$}
		
	\end{algorithmic}
	\caption{$ParaLiNGAM~Code patch$} 
	\label{alg:SchedulerV2}
\end{algorithm}

The description of $GPUKernel$ is given in Algorithm \ref{alg:GPUKernel}.
The main part of the algorithm is a loop in lines $1-13$. A worker's task is completed if it compares itself to all other workers, which is checked in line $13$.

In line $2$, same as line $14$ of Algorithm \ref{alg:ParaLiNGAM_FR}, workers check for their messages. Since GPU has limited resources, it makes a queue for blocks if it cannot fit them on its streaming multiprocessors (SMs). Hence, some workers launch later, therefore, checking messages as the first task would help them gain information from other previously active workers and check their scores to see if they have already reached the threshold.

In lines $3-5$, workers check their scores to see if they reached the threshold. In that case, they have to stop for this step and wait for the threshold update.
Then, in lines $6-8$, workers also check for termination of the iteration, and if they have finished their comparisons after checking messages, they set $finish$ as $True$.

Next, in lines $9-11$, workers choose their comparison target by increasing $\mathcal{C}[w]$.
Then, workers perform their comparisons in line $12$. Finally, if a worker finishes comparisons without reaching the threshold, it sets $finish$ as $True$ and notifies the host scheduler that  is the end of the current iteration.

In line $22$ in Algorithm \ref{alg:Scheduler}, we used a condition to avoid redundant comparisons. In discussed implementation on GPU, we perform comparisons asynchronously. Hence, checking redundant comparisons is not straightforward as before, and we need to use more complicated mechanisms. For brevity, such details are not mentioned in the algorithms. But, in short, for this matter, we utilize a flag for each comparison, and workers try to lock them with atomicCAS operation.
Moreover, if a worker reaches some comparison that its flag is already set, the worker skips that comparison and would receive that comparison's result later.

\begin{algorithm}[h]
	\begin{algorithmic}[1]
		\Input $State$
		\Output $finish$
		\Block p
		\Repeat
			\State $\mathcal{S}[w] += CheckMessages(w, State)$\hfill //Check
			\If {$\mathcal{S}[w] > \gamma$}\hfill //Evaluate
				\State \textbf{Exit}
			\EndIf
			\If {$\mathcal{D}[w, :]$ is all $True$}
				\State \textbf{break}
			\EndIf
			\Repeat \hfill // Compare
				\State $\mathcal{C}[w] = (\mathcal{C}[w]+1)\%size(U)$
			\Until{$\mathcal{D}[w][\mathcal{C}[w]]$ is $False$}
			\State $\mathcal{S}[w] += Compare(w, \mathcal{C}[w], state)
			\hfill$
		\Until{$\mathcal{D}[w, :]$ is all $True$}
		\State $finish = True$
	\end{algorithmic}
	\caption{$GPU Kernel$} 
	\label{alg:GPUKernel}
\end{algorithm}

\section{Experimental Evaluation}
\label{sec:exp}

\subsection{Setup}
\label{sec:exp:setting}

The proposed ParaLiNGAM algorithm is implemented in C++ language using CUDA parallel programming framework. The source code is available online \cite{sourceParaLiNGAM}. 

We experimentally evaluate ParaLiNGAM, along with DirectLiNGAM \cite{shimizu2011directlingam} which is a sequential method. The latest implementation of DirectLiNGAM is in Python language \cite{sourceDirectLiNGAM}. In order to have fair comparisons, we re-implemented DirectLiNGAM in C++ language. This implementation is available in \cite{sourceParaLiNGAM}. 


We employ a server machine with Intel Xeon CPU with $16$ cores operating at $2.1$ GHz. 
Since DirectLiNGAM is a sequential method, it is executed on a single core.
The CUDA kernels in ParaLiNGAM are executed on Nvidia Tesla V$100$ GPU, and the other procedures are executed sequentially on a single core. We use Ubuntu OS 20.04, GCC version 9.3, and CUDA version 11.1.

\subsection{Real-World Datasets}
\label{sec:exp:real_data}

\begin{table}[tp]
	\centering
	\caption{Benchmark datasets.}
	\label{tab:datasetSpec}
	\begin{tabular}{|c|c|c|}
	\hline
	Dataset 		& \# of reactions	& \# of non-zero variables ($p$)  \\
	\hline
	iML1515 			& 2712 							& 2326						\\
	\hline
	iEC1372\_W3110 			& 2758 						& 2339						\\
	\hline		
	iECDH10B\_1368 		& 2742 							& 2252						\\
	\hline
	iY75\_1357 & 2759 							& 2249						\\
	\hline
	iAF1260b 		& 2388 							& 1588					   \\
	\hline
	iAF1260 	& 2382 							& 1633						\\
	\hline
	iJR904 & 1075 							& 770						\\
	\hline
	E.coli Core &
	95								&	 85 \\
	\hline
	\end{tabular}
\end{table} 

We utilize seven Genome-scale metabolic networks as our benchmarks \cite{feist2007genome, reed2003expanded, monk2017ml1515, orth2011comprehensive, monk2016multi, monk2013genome, feist2010model} to evaluate the performance of ParaLiNGAM.
The metabolites are molecules involved in chemical reactions in a cell, and sets of these chemical reactions are so-called Metabolic networks \cite{lacroix2008introduction}. 
These metabolic networks can be studied in silico with  flux balance analysis \cite{orth2010flux} which is a way to simulate metabolic networks and to measure effects on the system by external
influences. The flux balance analysis is utilized to generate datasets.
The data generation procedure is performed with COBRA-toolbox \cite{becker2007quantitative, schellenberger2011quantitative}, and it considers the metabolic network of the Escherichia coli bacteria str. K-12. In order to utilize these networks, we acquire their models from BiGG Models\cite{king2016bigg}.

To generate data, first, we import models from BiGG models in COBRA. Then, utilizing COBRA's optGpSampler \cite{megchelenbrink2014optgpsampler}, we generate data by uniformly sampling from solution space with hit and run algorithm.
In this algorithm, first, $n$ points (samples) are generated in the middle of the solution, and then these points are relocated in a random direction. Although, after substantial steps in this procedure, some reactions' samples remain zero. These reactions are removed from datasets.

Details of the generated datasets are shown in Table ~\ref{tab:datasetSpec}. The first column shows the number of reactions, and the second column shows the number of non-zero variables among these reactions.
Each variable (reaction) is generated with $10000$ samples.

\subsection{Performance Comparison}

\subsubsection*{Comparing ParaLiNGAM with DirectLiNGAM }

The runtimes of both DirectLiNGAM and ParaLiNGAM are reported in Table \ref{tab:realTimings}.  The accuracy of the proposed solution is exactly the same as the DirectLiNGAM. Thus, we only report runtime of these experiments.
The third column shows serial runtime. The serial runtime on iJR904 dataset is 287780 seconds (approximately 3.3 days). 
Since the computational complexity has cubic relation with respect to the number of variables, the runtime of iAF1260b dataset, which is the next smallest dataset after iJR094, would probably be longer than three weeks. Thus, it is impractical to measure the runtime for other datasets, and the serial runtime is reported just for two datasets, which is 485 seconds for E.coli Core and 287780 seconds for iJR904. More comparisons with the serial solution is reported in Section \ref{sec:scalability} for synthetic datasets.

The fourth column reports runtime of ParaLiNGAM for all datasets, which ranges from 759 milliseconds to 91.3 seconds. The speedup ratio over serial execution for measured datasets is up to 3152.

\begin{table*}[tp]
	\centering
	\caption{Comparing the serial and parallel implementations. The third and fourth columns show the runtimes. The last column shows the speedup ratio, which is calculated by dividing the serial runtime over the parallel runtime.}
	\label{tab:realTimings}
	\begin{tabular}{|c|c|c|c|c|c|}
	\hline
	Dataset 		& \# of variables	& Serial runtime (sec.)  & GPU runtime (sec.) & Speedup ratio \\
	\hline
	iML1515 		& 2326	 & \textemdash \textemdash	& 1321			 & \textemdash \textemdash\\
	\hline
	iEC1372\_W3110	& 2339	 & \textemdash \textemdash	& 1420			& \textemdash \textemdash \\
	\hline		
	iECDH10B\_1368 	& 2252	&  \textemdash \textemdash & 1216			& \textemdash \textemdash \\
	\hline
	iY75\_1357 		& 2249	&  \textemdash \textemdash & 1174			& \textemdash \textemdash \\
	\hline
	iAF1260b 		& 1588	&  \textemdash \textemdash & 507			& \textemdash \textemdash \\
	\hline
	iAF1260 		& 1633	&  \textemdash \textemdash & 518			& \textemdash \textemdash \\
	\hline
	iJR904 			& 770	& 287780 ($\sim$3.3 days)	& 91.3	& 3152			\\
	\hline
	E.coli core 	& 85	& 485		& 0.759	& 638\\
	\hline
	
	\end{tabular}
\end{table*}
\subsubsection*{Comparing ParaLiNGAM with Other Parallel Methods}

\begin{figure*}[tp]
\centering
\includegraphics[width = \linewidth]{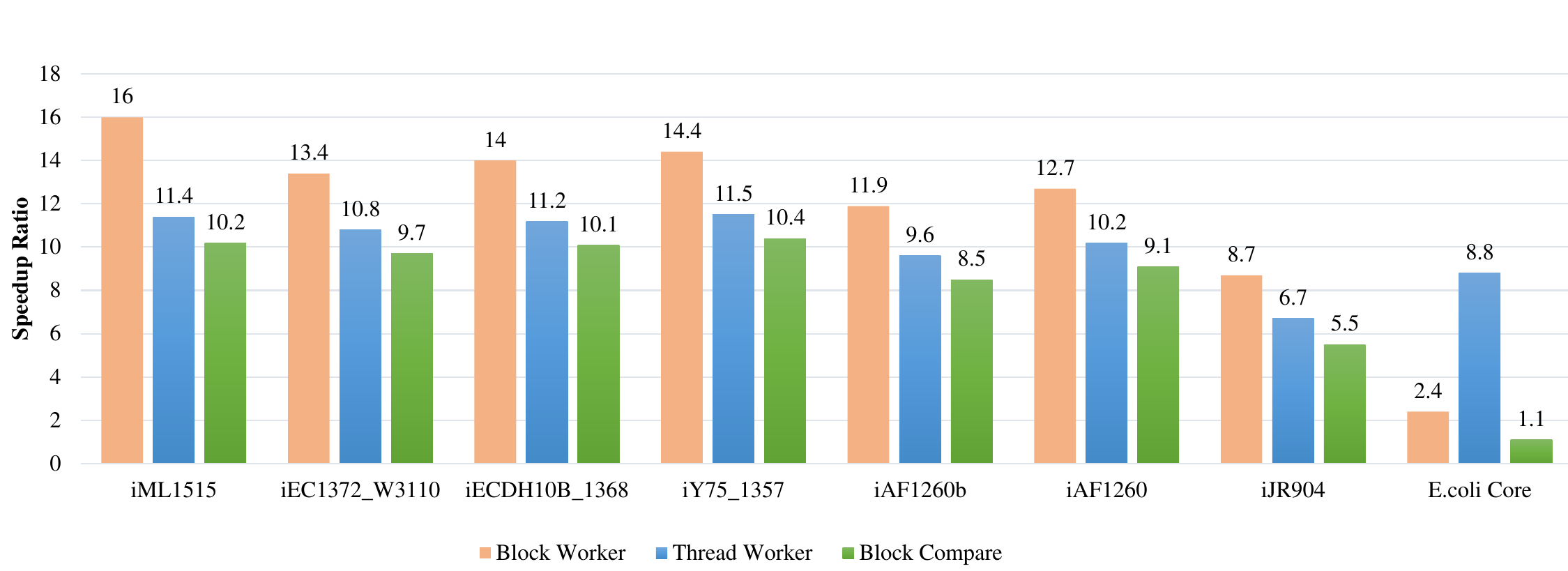}
\vskip -3mm
\caption{Comparing the performance of  ParaLiNGAM with three baseline algorithms on GPU. Every bar illustrates speedup ratio between the base line algorithm and ParaLiNGAM.}
\label{fig:baseLineTiming}	
\end{figure*}

Herein, three baseline parallel algorithms are introduced, and their performance are compared against the proposed ParaLiNGAM algorithm. See Fig. \ref{fig:baseLineTiming}. The first baseline algorithm is formed by assigning each variable to a block, and blocks compare themselves to each other. In specific, we have blocks equal to the number of remaining variables in each iteration, and each block performs one comparison at a time. We call this algorithm \textbf{Block Worker}.

The second baseline algorithm is similar to the previous one, in that  each variable is assigned to a block, but in each block, threads are responsible for different comparisons. Hence, each block can perform comparisons simultaneously. Since running lots of comparisons in parallel requires a lot of memory, this method cannot fit in GPU for large numbers of variables. In specific, we need separate memory for normalized variables and calculated residuals (see Compare Function in Algorithm \ref{alg:baseLiNGAM_FR}). As a result, we need $O(r^2 n)$ memory ($r$ is the number of remaining variables) in each iteration, which is problematic for large datasets. However, in Block Worker, we have just $r$ simultaneous comparison at a time and require just $O(r n)$ memory, which is a moderate value. 
Hence, we need to optimize memory usage in this algorithm. 
In specific, we store parameters like mean, variance, and covariance of comparing variables used in calculations. Then, we merely read input data from memory to perform comparisons. This solution not only solves memory issues but also reduces the runtime by avoiding redundant memory reads and writes. We call this algorithm \textbf{Thread Worker}.

In the third baseline algorithm, in each iteration, every comparison is assigned to a block, i.e., we have $r\times r$ blocks ($r$ is the number of remaining variables), and block of index $(i, j)$ compares $X_i$ with $X_j$. Like the previous baseline algorithm, performing many comparisons in parallel requires a lot of memory. Therefore, the above  mentioned optimizations are utilized to reduce memory usage in this algorithm as well. We call this algorithm \textbf{Block Compare}.



As illustrated in Fig. \ref{fig:baseLineTiming}, ParaLiNGAM  is 16 X to 2.4 X faster than the Block Worker method, 11.5 X to 6.7 X faster than Thread Worker, and 10.4 X to 1.1 X faster than Block Compare. Note that E.coli Core dataset has fewer variables compared to other datasets, and it under-utilizes the GPU. 
In most cases, Block Worker has the worst performance among the three baseline methods. This is due to its inefficient memory usage. Between Thread Worker and Block Compare, the first method employs too much parallelism, and threads of each block try to access  different parts of memory, which is less efficient compared with Block Compare, in which, there is a limited number of concurrent comparisons (number of blocks that can run simultaneously on the GPU). 

\subsection{Scalability}
\label{sec:scalability}
In this section, we evaluate the scalability of ParaLiNGAM. In particular, we measure the runtime of our proposed algorithm against DirectLiNGAM for different numbers of variables ($p$) and samples ($n$). 
 
We follow a similar procedure as ICA-LiNGAM \cite{shimizu2006linear} for the data generation mechanism.
First, we choose the number of parents for each variable, and generate a random matrix for adjacency matrix $B$. In sparse graphs, the number of parents is uniformly selected from interval $[1, 0.2 p]$, and for dense graphs the interval is $[0.25 p, 0.5p]$ ($p$ is number of variables).
Next, non-zero entries of $B$ are replaced with a random value, from interval $[-0.5, -0.95] \bigcup [0.5, 0.95]$.
Next, we generate exogenous noise $N_i$ for each variable by sampling from Gaussian distribution, then pass them through a power non-linearity (keeping same sign, but changing the absolute value to an exponent in interval $[0.5, 0.8] \bigcup [1.2, 2]$).
Finally, we generate samples for all variables recursively and permute them randomly. For this section, datasets are generated for number of variables of $p = 100, 200, 500, 1000$ and sample sizes of $n = 1024, 2048, 4096, 8192$.



\begin{figure*}[tp]
\centering
\includegraphics[width = 1.05 \linewidth]{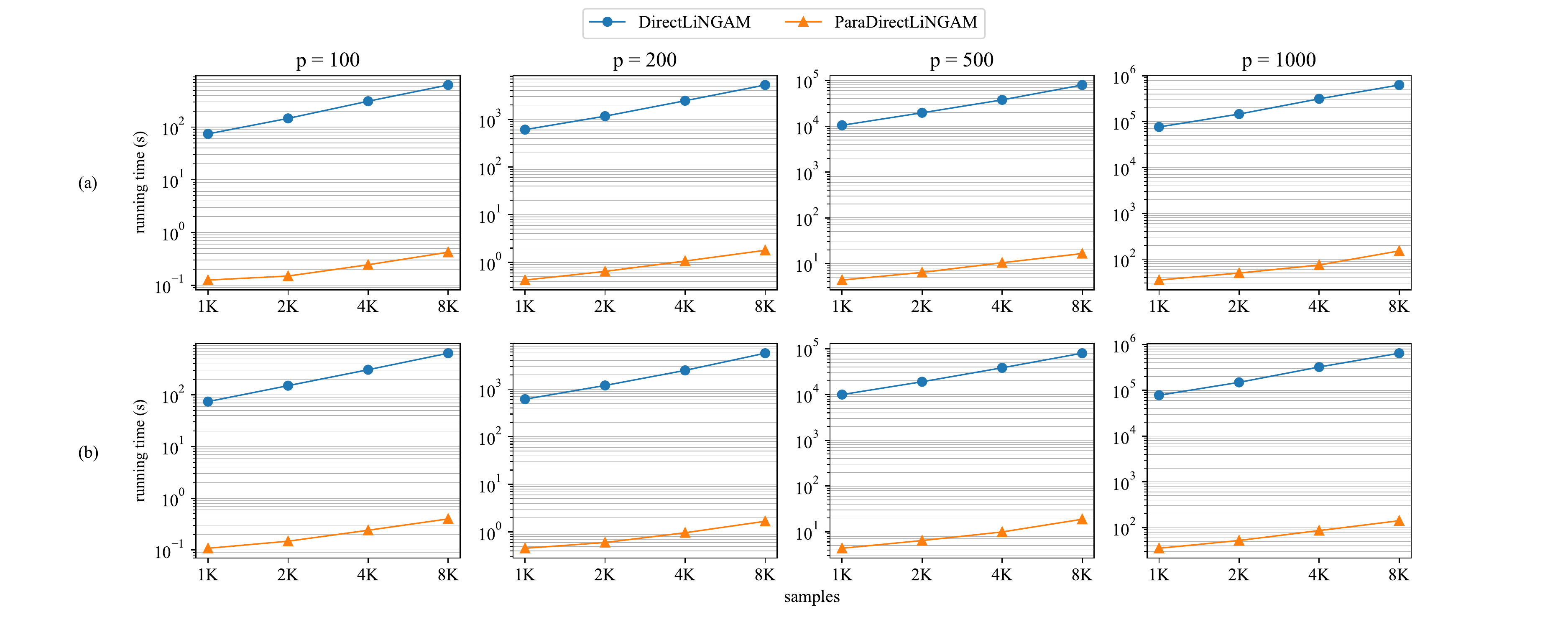}
\caption{Runtimes of ParaLiNGAM  and DirectLiNGAM for  a) sparse graphs and b) dense graphs for different number of variables and sample sizes.}
\label{fig:syntheticTiming}	
\end{figure*}

As mentioned earlier, the proposed solution does not change the algorithm accuracy and has the same precision. Therefore, only the runtime of the algorithms are shown in Fig. \ref{fig:syntheticTiming} for sparse and dense graphs.

In Fig. \ref{fig:syntheticTiming}, each column shows runtime for $p=100, 200, 500, 1000$, and rows (a) and (b) show runtime for sparse and dense graphs, respectively. The runtimes of sparse graphs are similar to dense graphs for DirectLiNGAM algorithm, due to the same procedure and computations. In particular,  the independent test is performed for all pairs of variables in all iterations despite the graph density. Hence, runtime merely depends on the number of variables and samples. The runtime of DirectLiNGAM varies from 71.4 seconds to 658806 seconds ($\sim$ 7.6 days). However, ParaLiNGAM attains a much smaller runtime compared with DirectLiNGAM, and its runtime varies from 119 milliseconds to 151 seconds. The speedup ratio of the proposed algorithm over the serial implementation ranges from 536 X to 4657 X. Furthermore, the speedup ratio increases as $p$ and $n$ increase.

\section{Conclusion}
\label{sec:conc}
In this paper, we proposed a
parallel algorithm for learning causal sturctures in LiNGAM model based on DirectLiNGAM algorithm. In the proposed algorithm, we employed a threshold mechanism to save a large portion of comparison in DirectLiNGAM. Moreover, we proposed a message mechanism and mathematical simplifications to further reduce the runtimes. Experiments showed the scalability of
our prospered algorithms with respect to the number of variables, the number of samples, and different types of graphs and achieved remarkable performance with respect to serial solution.

\appendices


%

\ifCLASSOPTIONcaptionsoff
  \newpage
\fi



%

%

 \bibliographystyle{IEEEtran}
\bibliography{reference}

\begin{thebibliography}{10}
\bibitem{sourceParaLiNGAM}
The source code is currently available only for the review process, but it will
  be publicly available on GitHub upon publication of the paper.
\end{thebibliography}

\newpage
 
\begin{IEEEbiography}[{\includegraphics[width=1in,height=1.25in,clip,keepaspectratio]{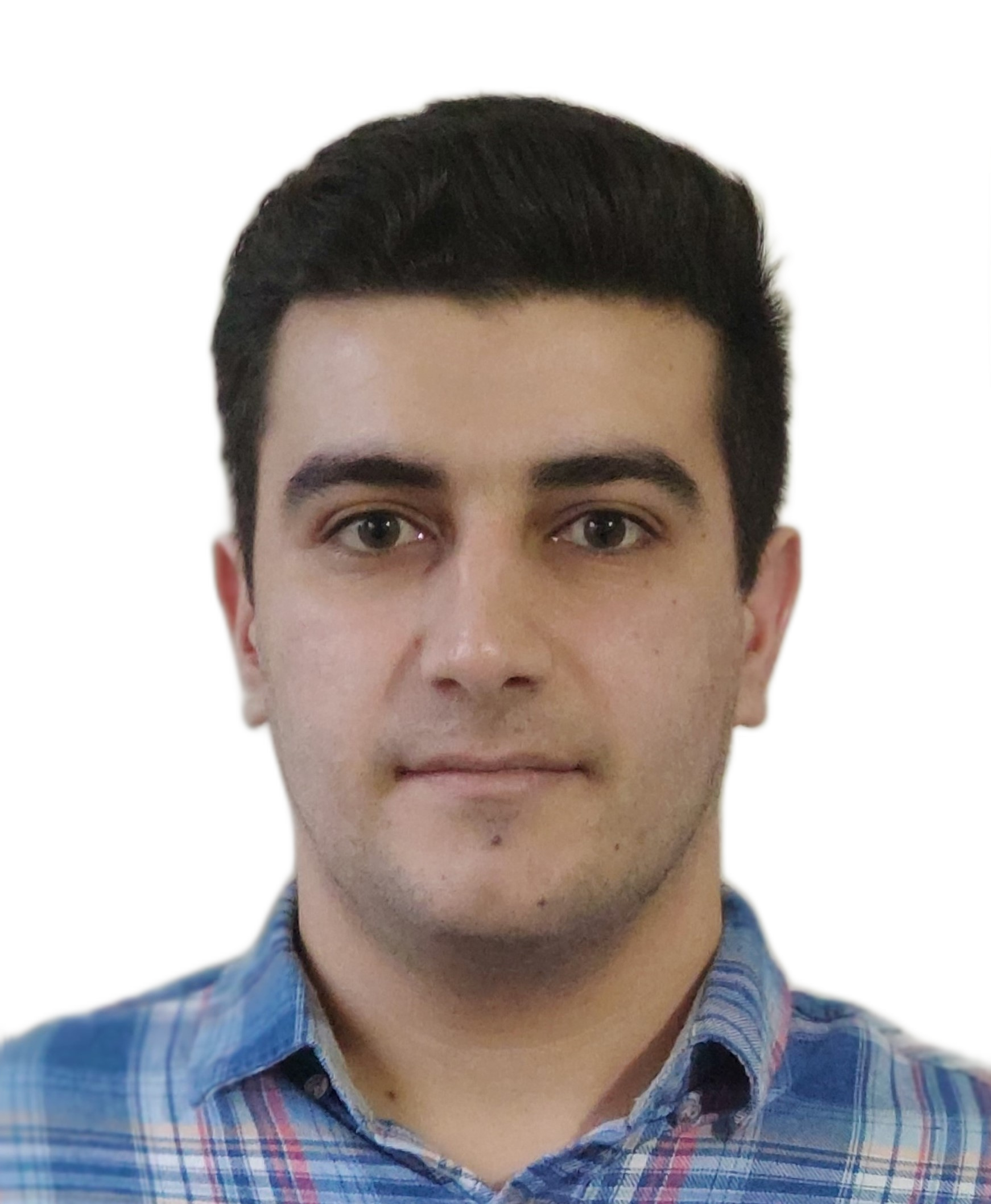}}] {Amirhossein Shahbazinia} received the B.Sc. degree in electrical engineering from University of Tehran, Tehran, Iran, in 2019. He is currently working towards the M.Sc. degree in electrical engineering at Sharif University of Technology, Tehran, Iran. His research interests include parallel processing, machine learning, and graphical model learning. 
\end{IEEEbiography} 

\begin{IEEEbiography}[{\includegraphics[width=1in,height=1.25in,clip,keepaspectratio]{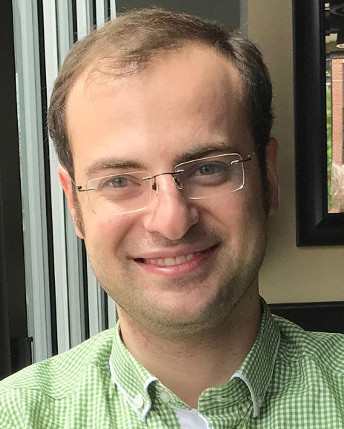}}] {Saber Salehkaleybar} received the B.Sc., M.Sc. and Ph.D. degrees in electrical engineering from Sharif University of Technology, Tehran, Iran, in 2009, 2011, and 2015, respectively. He is currently an assistant professor of electrical engineering at Sharif University of Technology, Tehran, Iran. His research interests include distributed systems, machine learning, and causal inference.
\end{IEEEbiography}

\begin{IEEEbiography}[{\includegraphics[width=1in,height=1.25in,clip,keepaspectratio]{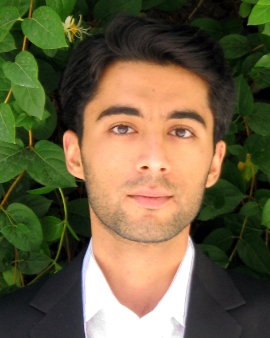}}] {Matin Hashemi} received the B.Sc. degree in electrical engineering from Sharif University of Technology, Tehran, Iran, in 2005, and the M.Sc. and Ph.D. degrees in computer engineering from University of California, Davis, in 2008 and 2011, respectively. He is currently an assistant professor of electrical engineering at Sharif University of Technology, Tehran, Iran. His research interests include algorithm design and hardware acceleration for machine learning and big data applications.
\end{IEEEbiography} 




\end{document}